\documentclass{aa}
\usepackage{graphics}
\begin{document}
\thesaurus{02(12.07.1; 11.17.4: QSO 0957+561)}
\title{Constraints on source and lens parameters from microlensing 
variability in QSO 0957+561 A,B}

\author{ S. Refsdal\inst{1,5}, R. Stabell\inst{2,5},
  J. Pelt\inst{3,5}, \and R. Schild\inst{4,5}}
\institute{Hamburger Sternwarte, Gojenbergsweg 112, D-21029 Hamburg,
 Germany
\and
 Institute of Theoretical Astrophysics, University of Oslo, 
 P.O. Box 1029, Blindern, N-0315 Oslo, Norway
\and 
 Tartu Observatory, 61602 Toravere, Estonia
\and
 Harvard-Smithsonian Center for Astrophysics, MS-19 60 Garden Street
 Cambridge, MA 02138, USA
\and 
 Centre for Advanced Study, Drammensveien 78, N-0271, Oslo, Norway}
\date{Received 28.02.2000, accepted ....}
\offprints{rolf.stabell@astro.uio.no}
\authorrunning
\titlerunning
\maketitle
\markboth{S. Refsdal et al. : Constraints from microlensing in QSO 0957+561}
         {S. Refsdal et al. : Constraints from microlensing in QSO 0957+561}
\begin{abstract}
From regular monitoring of the Double Quasar QSO 0957+561 A,B there is 
now general agreement on a time delay of about 416 days. This has made
it possible to determine the microlensing residual  in the light-curve,
see Pelt et al. (1998). We have used two significant microlensing 
features: 1) A ``quiet'' period with a variability less than 
0.05 mag lasting about 8 years, and 2) A change in the residual 
of 0.25 mag during a time interval of about 5 years.
The first feature gives a lower limit for the lens mass, $M$, for a given 
normalized source radius, $r$, whereas the second feature gives an
upper limit. We have considered the amount of mass in a continuum to
be a free parameter with possible values between 0\% (all mass in
lenses) and 90\%.
At a significance level of 1\% the mass can only be 
constrained within a rather wide range ( $10^{-6}M_{\odot}$ to
$5M_{\odot}$). For the radius $R$ of the source an upper limit of 
$10^{16}$ cm is found, whereas the normalized source radius $r$ is
restricted to be smaller than 30.
At a level of 10\% , however, the range of possible masses is much
narrower ($2\cdot 10^{-3} M_{\odot}$ to $0.5 M_{\odot}$), and the 
upper limit of $R$ is about $6\cdot 10^{15}$ cm, whereas the value of
$r$ is restricted to be less than 2. We have used an 
effective transverse velocity $V$ equal to 600 km s$^{-1}$. 
\end{abstract}

\section{Introduction}
    
The double quasar Q0957+561 A,B was the first discovered multiple image
gravitational lens, the first to have a measured time delay and
to produce a gravitational lens determination of the Hubble parameter, and
the first system in which a microlensing effect was seen in the observational 
brightness record (Vanderriest et al. 1989). This was not
unexpected since such microlensing effects had already been predicted
by Chang \& Refsdal (1979), see also Kayser et al. (1986), hereafter
referred to as KRS, and 
Schneider \& Weiss (1987). With the time delay confirmed by the
Vanderriest et al. report, Schild \& Smith (1991) noted evidence for 
fine structure in the microlensing light-curve. In this
paper we do not concern ourselves with this reported fine structure, and
focus instead on the long-term microlensing trends.

After a period of controversy about the correct time delay value, 
Kundi\'{c} et al. (1996) observed an
unusually distinct event in the light curve of image A which repeated
in image B 417 days later. This time delay was
very close to the values obtained earlier by Vanderriest et al. (1989),
Schild \& Smith (1991) and Pelt et al. (1996), and has later
been confirmed with high precision by other workers, see for instance 
Pelt et al. (1998) who found a time delay of 416.3 days.
A precise value of the time delay is necessary in order to subtract
out the intrinsic quasar brightness fluctuations on all time scales, 
and to find the microlensing residual.

The most accurate determination of microlensing variability in
Q0957+561 was reported by Pelt et al. (1998). This variability is shown in
Fig. 1 (their Fig. 9). Two principal 
features are seen in the 15 year microlensing light-curve: a rise 
of 0.25 mag during a period of about 5 years
(1982-1986) with a maximum slope of 0.07 mag per year, and a quiet
phase of about 8 years (1988-1996) with a variability less than 0.05 
mag. The original data in Pelt et al. (1998) show a large scatter which
has sometimes been suggested to evidence microlensing on time scales of
100 days and shorter (Schild 1996) but the existence of such low amplitude
and rapid fluctuations is complicated by questions of the accuracy of the
data; we therefore restrict our analysis to the less controversial
long-term microlensing. Although gaps as long as 300 days are present in
the long-term microlensing record, segments where the brightness record 
is more intensively sampled seem to show that amplitudes on sub-year
timescales are less than 5\% and they average away as shown convincingly in
the Pelt et al. (1998) Fig. 9 plot.
 
This paper is concerned with understanding the two
features in the light-curve mentioned above, from models of the 
gravitational lens and 
microlensing from stars or other compact objects presumed to lie 
within the lens galaxy G1. For typical quasar accretion disc sizes 
it is possible that statistical effects from numerous stars projected in
front of the quasar must be considered; a statistical
microlensing theory for such a case can be found in Refsdal \&
Stabell (1991 (RS1), 1993 (RS2) and 1997). 

From macro-lens modeling 
of Q0957+561 one finds the lensing optical depth
for the A and B image to be $\kappa_{\rm A}=0.22$ and $ \kappa_{\rm B}=1.24$,
respectively, and with shear terms $\gamma_{\rm A}=\pm 0.17$ and
$\gamma_{\rm B}=\pm 0.9$ (Lehar, private communication; see also Schmidt \&
Wambsganss (1998) for similar estimates and a definition of the
shear). Our source model is
circular with a Gaussian distributed surface luminosity and ``radius''
$R$. 

We shall treat three
different mass models in our microlensing simulations:\\
Case 1: All the lensing mass is in
identical compact micro-lenses with mass $M$.\\
Case 2: 10\% of the lensing mass is in
identical compact lenses with mass $M$ and 90\% in an evenly distributed
continuum.\\ 
Case 3: 10\% of the lensing mass is in compact lenses with $1~M_{\odot}$ 
and 90\% in compact lenses with mass $M < M_{\odot}$.

Our approach is to compare simulated microlensing light-curves 
with the observed microlensing variability. 
Because the lens mass $M$ and the source radius $R$ both
affect the amplitude and time scale of variability, we find some
interesting constraints for these two parameters in the lens 
system Q0957+561. 

Schmidt \& Wambsganss (1998)
have examined the available parameter space based upon an observed 
quiet period of 160 days with a variability less than 0.05 mag.

The results of our study are based on the microlensing variability 
during 15 years, including a long quiet phase as well as a 
significant event and naturally gives stronger constraints.

\section{Observed microlensing variability}

We make use of the observed microlensing light-curve
covering a time span $T=15$ years, see Fig.1. The curve is given by:

\begin{figure}[tb]
\resizebox{8.5cm}{!}{\includegraphics{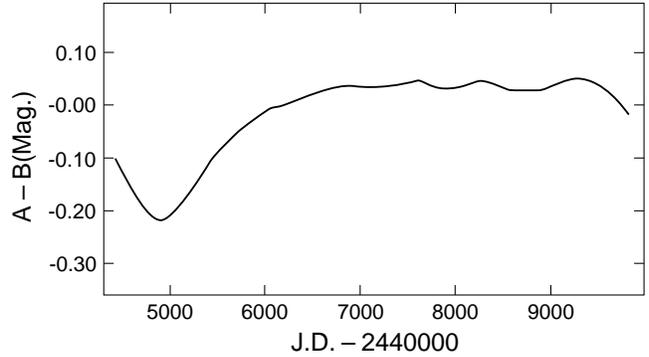}}
\caption[]{Observed microlensing light-curve for QSO 0957+561, 
m(A) minus the timeshifted m(B),
showing an event of 0.25 mag lasting 5 years and a quiet phase with 
a variability $\delta m$  less than 0.05 lasting 8 years}

\end{figure}

\begin{equation}
m(t)=m_{\rm A}-m_{\rm B}^{\tau}
\end{equation}

\noindent
where $m_{\rm B}^{\tau}$ is
the magnitude  of the B image shifted by the time delay
$\tau=416.3$ days  (see Pelt et al. 1998).
Two significant microlensing features are found:\\
1) A ``quiet'' period with a variability
$\delta m=m_{\rm max}-m_{\rm min}$ less than $0.05$ mag lasting 
$\delta t=8$ years.\\
2) A variation of $\Delta m=0.25$ mag during a time
interval $\Delta t=5$ years, with a maximum slope of about 0.07 
mag/year (``the event'').

We shall treat these two features independently and investigate by
means of simulations how lens and source parameters can be
constrained. The results will be demonstrated in ``exclusion 
diagrams'', see Figs. 2 and 3.

Since the optical depth is much larger for the B image than for A,
the microlensing variability in the A image is likely to be
much smaller than in the B image. We shall therefore in this paper
neglect a possible microlensing variability in the A image and assume 
that the
observed microlensing variability comes only from B.
From test calculations we find that this approximation has a rather
small influence on the constraints obtained (see Discussion).

\section{Case 1: all mass in identical lenses}
We shall first choose all mass to be in lenses with mass $M$. This may
not be very realistic, but it allows a relatively simple discussion 
with only two free parameters ($M$ and source radius $R$), and
it gives us an understanding which is quite useful also for more
complex mass distributions. We simulated light-curves for the B image
with effective transverse motion of the source parallel 
to the shear and perpendicular to the shear
$(\gamma_{\rm B}=\pm 0.9)$. The results for these two cases are always
sufficiently similar so that an intermediate case easily can be
interpolated. For practical reasons we shall instead of $R$ use
the normalized source radius $r=R/R_{\rm E}$ as a free parameter,
where $R_{\rm E}$ is the Einstein radius (for mass $M$), projected into 
the source plane

\begin{equation}
R_{\rm E} =\sqrt{ \frac{4GM}{c^{2}}\frac{D_{\rm ds}D_{\rm s}}{D_{\rm d}}}=
4.8\cdot 10^{16}\sqrt{\frac{M}{M_{\odot} h_{60}}}~ \mbox{cm}
\end{equation} 

\noindent
Here the D's are angular size distances and $h_{60}$ is the
dimension-less Hubble parameter, $h_{60}=H_{0}/(60$ km s$^{-1}$
Mpc$^{-1})$. We have used a cosmological model with $\Omega =0.5$ 
and $\Lambda=0$.

For randomly distributed compact lenses we simulate light-curves for 
different
values of the normalized source radius $r$, using well known ray 
tracing techniques, see KRS and Schneider \& Weiss (1987). We
obtain for each value chosen for $r$ the magnitude of the source as a
function of the normalized source position: $m=m(y)$, where 
$y=\eta /R_{\rm E}$ and $\eta$ is the length coordinate along the source
track. For each value of $r$ we get the light-curve for different (but
still all identical) $M$-values by a transformation of $y$ to the time
$t$ measured by the observer. For our system we find

\begin{equation}
t = 25 y~v^{-1}_{600}\sqrt{\frac{M}{M_{\odot} h_{60}}}
~ \mbox{years} 
\end{equation}

\noindent
Here $v_{600}=V/(600$ km s$^{-1})$ where $V$ is the effective 
transverse velocity of the source 
assumed to be constant,  with time measured by the observer, see
KRS. When not stated otherwise we shall use $v_{600}=1$ and 
$h_{60}=1$ in this paper.

Since $t \propto y~\sqrt{M}$ the light-curves are simply
stretched out in time for larger values of $M$. Hence, it is not
necessary to calculate light-curves for different $M$-values (for
a given $r$). We have therefore
simulated light-curves only for various $r$-values ($r$=0.02, 
0.05, 0.1, 0.3, 1, 3, 5, 10, 20 and 50). The length of the 
light-curves corresponds in most cases to at least 1000 years for 
the allowed masses, see Eqs. (3) and (4). We first discuss the
constraints on $M$ and $R$ (and $r$) which can be obtained from the 
quiet phase of 8 years.

\smallskip
\noindent
i) Quiet phase constraints

\noindent
For each value of the normalized source radius $r$ we calculate
the microlensing light-curve $m(y)$ for $\kappa=\kappa_{\rm B}=1.24$ and 
$\gamma=\gamma_{\rm B}=\pm 0.9$. Giving equal weight to the two shear 
directions we can then
determine the probability $P_{\rm q}(M,r)$ to obtain a quiet period
lasting at least 8 years during a period of 15 years with a 
variability $\delta m = m_{\rm max} - m_{\rm min}$ less than $0.05$ mag. 
Since $t$ is proportional to
$y\cdot \sqrt{M}$, the probability $P_{\rm q}(M,r)$ 
obviously increases monotonically with
$M$. Hence, we can determine a critical mass $M_{\rm q}(r)$ by
requiring that $P_{\rm q}(M_{\rm q},r)$ is, say 0.01. We then choose to
exclude values of $M(r)$ less than $M_{\rm q}(r,P_{\rm q})$ since the
probability of getting a quiet period of 8 years or more is then less
than one percent (compare the paper by Schmidt \& Wambsganss, 1998).
The dependence of $M_{\rm q}$ on $r$ for $P_{\rm q}=0.01$ is shown in the 
exclusion diagram in Fig. 2 as the solid line to the left 
labeled {\bf q}. 

\begin{figure}[tb]
\resizebox{8.5cm}{!}{\includegraphics{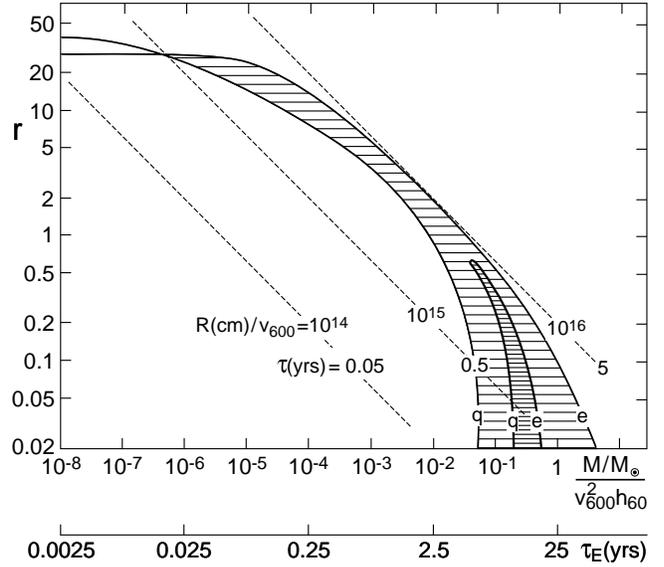}}
\caption[]{Exclusion diagram for the case when all mass is in lenses of
identical masses (Case 1). The constraint curves for the quiet phase
({\bf q}) and the event ({\bf e}) are shown in the
($M,r$)-plane. The allowed region for significance level 0.01 is
hatched and for significance level 0.1 more densely hatched. 
Note that the mass values scale with the effective
transverse velocity and the Hubble parameter.}

\end{figure}

We can roughly distinguish 3 different parts of the curve
$M_{\rm q}(r,0.01)$:

\noindent
1) $~r < 0.5$ : Here we find $M_{\rm q}$ very nearly constant and
equal to about $0.05M_{\odot}$. This corresponds to $R_{\rm E} \approx
10^{16}$ cm and $\tau_{\rm E}=R_{\rm E}/V \approx 5$ years.

The behavior with an almost constant $M_{\rm q}$ (and $R_{\rm E}$ 
and $\tau_{\rm E}$)
for small values of $r$ is a consequence of the nearly
identical light-curves for different (small) $r$ with the same star 
field. Only during caustic crossings will there be a major difference
in the light-curves, but then the variability will be large anyway and 
therefore violate the quiet phase condition. For small $r$ the
probability of getting a quiet phase therefore only depends on $M$ so that
$M_{\rm q}$ is nearly independent of $r$ for $r<0.5$.

\noindent
2) $~0.5<r<10$ : Here we find that $M_{q}$ is roughly
proportional to $r^{-2}$. Since $r=R/R_{\rm E}\propto
R/\sqrt{M}$ this means that $R$ is roughly constant in this
range. We find here $R\approx 4\cdot10^{15}$ cm and
$\tau=R/V\approx 2$ years. This result is not unexpected,
since the variability time scale for sources with $r>1$ is
about equal to $2 \tau$, see RS2.

\noindent
3) For $r>10$ the curve gets flatter and approaches
asymptotically a maximum value of $r_{\rm qm}\approx 40$. Such an
asymptotic behavior was to be expected from the results in RS1,
where it was shown that microlensing amplitudes for large sources
($r>>1$) are proportional to $r^{-1}$.

\smallskip
\noindent
ii) Event constraints

\noindent
In the simulated light-curves we consider only events with a
variability between $0.2$ mag and $0.3$ mag ($\Delta m \pm 0.05$), and
with a maximum slope during the event steeper than 
0.07 mag per year. The reason for choosing also an upper limit
for the allowed event magnitudes is that the simulated events for
small values of $r$ typically show larger amplitudes than the
observed one. In our analysis we have found it more practical to 
use the maximum slope rather than the time scale $\Delta t$. 
We can then estimate the probability $P_{\rm e}(M,r)$ of getting such
an event during a time span of $T=15$ years. Since the slope scales as
$M^{-0.5}$ and the simulated time span as $M^{0.5}$, it is clear that
$P_{e}(M,r)$ decreases monotonically with increasing $M$. We can
therefore determine a critical mass $M_{\rm e}(r)$ by requiring that
$P_{\rm e}(M_{\rm e},r)$ is, say $0.01$. 
We then choose to exclude values of
$M(r)$ larger than $M_{\rm e}(r,P_{\rm e})$.

The dependence of $M_{\rm e}$ on $r$ for $P_{\rm e}=0.01$
is shown in Fig. 2 as the solid line to the right labeled {\bf e}.
It is seen that $M_{\rm e}(r)$ is roughly proportional to $r^{-2}$,
and therefore $R$ is fairly constant over a large range of $r$-values
($0.2<r<10$). We get here $R\approx 10^{16}$ cm and
$\tau=R/V=5$ years. This is easy to understand since a typical
event time scale is the crossing time $2R/V$ for large sources (see
RS2) as well as for small sources (caustic crossings).

For small $r$-values we see that the $M_{\rm e}$-curve departs from
$R\approx $ constant towards smaller values of $R$, approaching a
constant $M_{\rm e}$-value of about 5$M_{\odot}$ for very small $r$. 
The reason is that
for such small $r$-values, caustic crossing events as
 small as $\Delta m \approx 0.25$ are quite rare.
Of some importance here are, however, some non-caustic events, 
typically occurring 
when the track passes just outside a cusp of a caustic. For very small 
sources($r<0.1$) these events are nearly independent of $r$, which can
explain that $M_{\rm e}(r)$ stays nearly constant for these small
$r$-values. 
We note that the time scale
for these events are usually longer than the crossing time $2R/V$, see
also Wyithe et al. (1999).

For very small $r$-values the effect of the so-called ghost caustics
which are produced when two lens masses are very close to each other
is also of importance, (Grieger et al. 1989, Wambsganss \& Kundi\'{c}
1995). These weak caustics can often
``produce'' events during caustic crossings as small as about 0.25 mag, 
even for very small
$r$-values. An uncertainty arises, however, because relative motion of
the stars some times causes large velocities of the ghost caustics so 
that the ``ghost events'' will occur more often and on a
shorter time scale. Since we have not considered relative motion in our
calculations, we may have underestimated the frequency and
overestimated the time scale of ghost
events. In that case the upper limit for $M_{\rm e}$ has been
underestimated. 

For large $r$-values the $M_{\rm e}$-curve also departs from $R\approx $
constant towards smaller values of $R$ and approaches asymptotically
a maximum $r$-value of about $r_{\rm em} = 30$. This high value of
$r_{\rm em}$ were to be expected from the results in RS2, where it was
shown that quite significant events can be produced for rather large
values of $r$.

\smallskip
\noindent
iii) The allowed region

\noindent
From the discussion in section i) we found that the quiet phase
constraint gives a lower limit, $M_{\rm q}(r)$ for the lens mass. 
In section ii) on the other hand, the event 
constraint gave an upper limit
$M_{\rm e}(r)$. Hence, at a significance level of $P$ we are left with an
allowed region between the $M_{\rm e}$ and $M_{\rm q}$-curves in 
the $(M,r)$-plane which fulfills

\begin{equation}
M_{\rm q}(r,P)<M(r)<M_{\rm e}(r,P)
\end{equation} 

\noindent
For $P=0.01$ the allowed region is shown hatched in Fig. 2.
For $r$ between 0.5 and 20
we find a strip within a relatively narrow range of 
$R$-values ($4\cdot 10^{15}$ cm$<R<10^{16}$ cm ), covering 
$M$-values between $10^{-6}M_{\odot}$ and
$10^{-1}M_{\odot}$. For $r<0.5$, mass values between
$3\cdot 10^{-2}M_{\odot}$ and $1M_{\odot}$ are usually allowed. 
For the smallest $r$-values the upper limit reaches about 5$M_{\odot}$.
We cannot give a reliable lower limit of $r$. However, the lack of any clear
high magnification events in all observed quasar light-curves is an
argument against too small $r$-values (and $R$-values) for QSOs in 
general.

For larger $P$-values the allowed
region obviously decreases. As an example the allowed region for 
$P=0.1$ is shown more densely hatched in Fig. 2.
It is interesting to note that Eq. (4) can now  only be fulfilled 
for $r<0.6$ and the possible mass shrinks approximately to the 
interval between 0.05 $M_{\odot}$ and 0.5 $M_{\odot}$ (significance level of
10\%, and all mass in identical lenses).

\section{Case 2: 90\% of the lensing mass in a continuum}
Since a large fraction of the lens mass may be in a smoothly
distributed continuum, we have repeated most of the calculations in
the previous section, but now with 90\% of the mass in a continuum and
the remaining 10\% in compact lenses with mass $M$. The resulting
constraint curves are shown as solid lines in Fig. 3 for a
significance level of 0.01, and the allowed region is hatched. 
We find that the asymptotic maximum $r$-values $r_{\rm qm}$ and 
$r_{\rm em}$ decrease
with a factor of about 3 relative to Case 1.
This was to be expected because the variability
amplitude for large $r$ is proportional to
$\sqrt{\kappa_{\rm s}}/r$, so that $r_{\rm qm}$ and $r_{\rm em}$ should be 
proportional to $\sqrt{\kappa_{\rm s}}$, see RS1.
Here $\kappa_{\rm s}$ is the optical depth for microlensing which is now
a factor of 10 smaller than in Case 1. 

\begin{figure}[tb]
\resizebox{8.5cm}{!}{\includegraphics{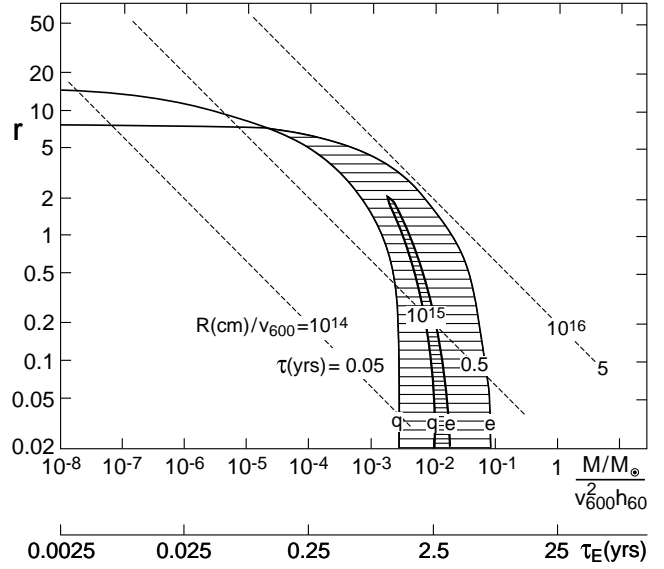}}
\caption[]{Same as Fig. 2, but with 90\% of the lens mass in a
continuum (Case 2)}
\end{figure}

The possible mass range is now restricted to an interval between 
$3\cdot 10^{-5}M_{\odot}$ and $10^{-1} M_{\odot}$.

For $P=0.1$ the allowed region is shown more densely hatched. We see
that the possible mass now only covers an interval between 
$2\cdot 10^{-3}M_{\odot}$ and $2\cdot 10^{-2} M_{\odot}$, and $r$ is
restricted to be less than about 2.

\section{Case 3: Two lens populations}
From the observed light and spectrum of the lensing galaxy we know
that some of the mass must be in stars with typically one solar mass.
We have therefore investigated a model with two lens populations; 
10\% of the mass in solar mass stars and the rest of the mass in 
objects of smaller mass, $M$. Treating as before $M$ and 
$r=R/R_{\rm E}(M)$ as free parameters we
get mostly small changes (relative to Case 1) in the  constraint curves
in the ($M,r$)-plane. The $M_{\rm q}$-curves are almost 
unchanged, since the few $1M_{\odot}$ stars can only
make a minor change in the probability of getting long quiet phases
for $M<1 M_{\odot}$.
The $M_{\rm e}$-curves also follow very closely
the corresponding curves for Case 1 (with all mass in identical
stars). A separate discussion of this case is therefore not necessary.

\section{Change of $V$ and $H_{0}$}
We have until now assumed specific values of the effective transverse 
velocity $V$ and the Hubble parameter $H_{0}$ ($v_{600}=1$ and $h_{60}=1$).
It is easy to see what happens if these values are changed. If for
example the transverse velocity is increased by a factor $F$, an
identical light-curve is obtained if the source size $R$ and the 
Einstein radius $R_{\rm E}$ are increased by the same factor, such 
that $r=R/R_{\rm E}$ is unchanged. Increasing the Einstein
radius by a factor F means that each lens mass is increased by a 
factor $F^{2}$, and to keep the optical depth for lensing unchanged, 
the linear size of the lens field must of course be scaled up with 
a factor $F$.

Correspondingly, an increase of the Hubble parameter with a factor $F$
can be compensated by an increase of each lens mass with the same
factor to keep the light-curve unchanged. Since the Einstein radius is
unchanged in this case, the optical depth is also unchanged, and no 
change in the size of the lens field is necessary.

According to the above discussion, the allowed regions in Figs. 2 
and 3 will be shifted in mass when $V$ and $H_{0}$ are changed. 

\begin{equation}
M_{\rm q}(r)\propto V^{2} H_{0}
\end{equation}

\begin{equation}
M_{\rm e}(r)\propto V^{2} H_{0}
\end{equation}

This is most conveniently taken into account by just scaling the 
mass coordinate appropriately,
thereby leaving the curves in Figs. 2 and 3 in the same 
positions. Obviously the dominating uncertainty in $M$ comes from 
the $V^{2}$ term.

\section{Discussion}
By means of the two features in the microlensing light-curve 
for the Q0957+561 system and numerical simulations, we have shown that it
is possible to set constraints on some parameters for the 
lens system. With the amount of the continuum mass between 0\% (Case
1) and 90\% (Case 2) we find the possible lens masses to have values
between $10^{-6} M_{\odot}$ and $5 M_{\odot}$ (1\% level) and between  
$2\cdot 10^{-3} M_{\odot}$ and $0.5 M_{\odot}$ (10\% level). The
maximum source radius is found to be $10^{16}$ cm (1\% level) and
$6\cdot 10^{15}$ cm (10\% level). By including Case 3 (giving
approximately the same reults as Case 1), we find only
negligible changes in the above limits.

From the two features in the microlensing light-curve 
it seems not possible to give a lower limit of $R$. The reason for
this is that the light-curves for very small (but different)
values of $r$ are very similar, except during caustic crossings, and
then the quiet phase constraint is violated anyhow.

From lens arguments based on the Q0957+561 light-curve 
alone, we can therefore not rule out very small
$r$-values (even those much smaller than 0.02), and hence very 
small $R$-values. However, the lack of any clear
high magnification events in observed quasar light-curves is an
argument against too small $r$-values (and $R$-values) for QSOs in 
general.

We have in this paper only considered microlensing effects in the B
image. The combined A and B microlensing light-curve will statistically
show slightly more variability and less chance of extended quiet
phases, and have a slightly higher rate of events than for B
alone. Hence, both $M_{\rm q}$ and $M_{\rm e}$ will increase slightly. Our
test calculations indicate a rather small effect of about 10\%.

Since we have neglected the relative motion of the lens masses we may
have underestimated the maximum allowed lens mass which occur for
small values of $r$, particularly in Case 1. This ought to be
investigated in the future.

We have in this investigation considered the two features found in the
light-curve for Q0957+561 independently and used the probabilities
$P_{\rm q}$ and $P_{\rm e}$ separately. An alternative approach, which would
have given much stronger constraints, might have been to consider the
probability  for both features to be found within a
period of 15 years which must be less than $P_{\rm q}\cdot P_{\rm e}$. 
The largest values of $P_{\rm q}\cdot P_{\rm e}$ (between 0.01 and 
$\approx 0.02)$ are
obviously found within the densely hatched regions in Fig. 2 (Case 1)
and Fig. 3 (Case 2). The most probable mass values would then mainly 
depend on the amount of mass in the continuum ($\approx 0.3 M_{\odot}$ in 
Case 1 and $\approx 0.01 M_{\odot}$ in Case 2), and the most probable
$r$-value is less than 1. It is, however, not 
possible to give levels of significance here since we are obviously 
doing a posteriori statistics.
It seems in fact difficult to avoid a posteriori statistics when
making use of a larger part of the information in the light-curve.

Wyithe et al. (1999) have published a series of papers (see astro-ph/9911245
and references therein) where they make use of the distribution of the
microlensed light-curve derivatives, and they apply the method on the
QSO 2237+0305 light-curve. Their method does not suffer from a posteriori 
statistics, but more information in the light-curve could perhaps have
been taken into account, for instance the correlation time-scale for 
the derivatives,
without running this risk. A comparison with their results is ,
however, difficult since they are considering a different system. For
their system the value of $R_{\rm E}$ is about a factor of 3 larger,
whereas the source radius $R$ is roughly a factor 2.5 smaller 
(estimated from the difference in intrinsic luminosity). Hence, the 
normalized source radius $r=R/R_{\rm E}$ will be a factor of about 7
smaller. With this factor in mind we do not see any serious conflict
between the rather small $R$-values they obtain and our (most
probable) values.

\end{document}